\documentclass[10pt]{IEEEtran}
\usepackage[ruled,vlined]{algorithm2e}
\usepackage{array}
\usepackage[caption=false,font=normalsize,labelfont=sf,textfont=sf]{subfig}
\usepackage{textcomp}
\usepackage{stfloats}
\usepackage{url}
\usepackage{verbatim}
\usepackage{graphicx}

\usepackage{color}
\usepackage[hidelinks]{hyperref}

\usepackage[numbers]{natbib}

\usepackage{balance}
\usepackage[dvipsnames]{xcolor}
\usepackage[many]{tcolorbox}
\usepackage{booktabs}

\usepackage{amsmath,amsfonts}
\usepackage{amssymb,amsthm}

\theoremstyle{definition}

\def\de{\overset{\Delta}{=}}

\SetKwInput{KwInput}{Input}
\SetKwInput{KwOutput}{Output}
\SetKwComment{Comment}{/*}{ */}

\title{Large Deviation Analysis of Score-based Hypothesis Testing}

\author{Enmao Diao, Taposh Banerjee, \emph{Member, IEEE}, and  Vahid Tarokh, \emph{Fellow, IEEE}
\thanks{Enmao Diao and Vahid Tarokh are with the Department of Electrical and Computer Engineering, Duke University, Durham, NC 27708, USA. Taposh Banerjee is with the Department of Industrial Engineering, University of  Pittsburgh, PA 15213, USA. Vahid Tarokh was supported in part by Air Force Research Lab grant number FA-8750-20-2-0504. Taposh Banerjee was supported in part by the U.S. Army Research Lab under grant W911NF2120295. \newline
}
}

\begin{document}
\maketitle
\begin{abstract}
Score-based statistical models play an important role in modern machine learning, statistics, and signal processing. For hypothesis testing, a score-based hypothesis test is proposed in \cite{wu2022score}.  We analyze the performance of this score-based hypothesis testing procedure and derive upper bounds on the probabilities of its Type I and II errors. We prove that the exponents of our error bounds are asymptotically (in the number of samples) tight for the case of simple null and alternative hypotheses.  We calculate these error exponents explicitly in specific cases and provide numerical studies for various other scenarios of interest.
\end{abstract} 


\section{Introduction}

Score matching~\cite{hyvarinen2005estimation, hyvarinen2007some} is a procedure that has emerged due to its ability to outperform likelihood-based benchmarks in image generations~\cite{song2020score, vahdat2021score}. It has been shown that it is possible to efficiently model $\nabla \log p(x)$ using deep neural networks (DNNs), where $p(x)$ denotes the data generating density. However, it is computationally non-trivial to calculate even un-normalized versions of $p(x)$ from these trained DNNs. For log-likelihood-based hypothesis testing and change detection, there exists a well-established literature. Recently change detection and hypothesis testing for score-based systems have been studied. In particular, score-based hypothesis testing \cite{wu2022score} was proposed as a procedure to decide whether or not to reject a hypothesis. We note that the likelihood ratio test (LRT) is a standard method commonly used for hypothesis testing, and the celebrated Neyman-Pearson lemma gives the uniformly most powerful optimality property of LRT for simple hypotheses testing. This implies that score-based hypothesis testing cannot outperform the LRT test for simple hypotheses testing when the densities of data under the null and alternative hypotheses are exactly known. However, the evaluation of exact likelihoods may be computationally cumbersome (or even intractable) for various modern statistical models including graphical models~\cite{graphic_models}, energy-based models~\cite{LeCun2006ATO} and deep generative models~\cite{Papamakarios2021NormalizingFF}. This gives some importance to score-based hypothesis testing \cite{wu2022score} and the analysis of its performance. This motivates our work in this paper, where we derive upper bounds on the probabilities of its Type I and II errors for score-based hypothesis testing, and establish that the exponents of our error bounds are asymptotically (in the number of samples) tight for the case of simple null and alternative hypotheses.

The outline of our paper is given next. In Section~\ref{sec:score_matching_test}, we will motivate the problem to be considered in this paper, review the related work, and present the score-based test for binary hypothesis testing \cite{wu2022score}. 
We then derive upper bounds on the probabilities of its Type I and II errors for finite sample size in Section~\ref{sec:FixedSampleError}. In Section~\ref{sec:LargeDeviation}, we show that these bounds are asymptotically tight using large deviation theory \cite{Cramer, dembo2009large}. In Section~\ref{sec:EstimatingErrExpo}, we discuss simulation methods for estimating these error exponents. In Section~\ref{sec:ExampleCalculations}, we provide explicit calculation for the error exponent in the multivariate Gaussian case. 
In Section~\ref{sec:NumericalResults},  we show the accuracy of these error bounds for experiments performed on real and simulated data. 
We will make our concluding remarks in Section~\ref{sec:conclusion}.

\section{Score-based Hypothesis Testing for Unnormalized Statistical Models}
\label{sec:score_matching_test}

\subsection{Binary Hypothesis Testing}
Let ${x}\in \mathcal{X}\subseteq\mathbb{R}^{d}$ be the realization of a random vector $X$. We denote $\mathbf{X}_n \de \{{X}_1, \dots,{X}_n\}$ as independent and identically distributed (i.i.d.) observations from an unknown distribution $P$ with probability density $p$. For two probability measures $P_\infty$ and $P_1$ with respective densities $p_\infty$ and $p_1$, 
we investigate the simple binary hypotheses testing problem:
\begin{align}
\label{eq:composite_hy}
   \mathcal{H}_0: p = p_\infty \quad \text{against} \quad \mathcal{H}_1: p =p_1.
\end{align}
It is well-known that the optimal test in Bayesian, minimax, and variational settings is the likelihood ratio test (LRT):
\begin{equation}
    \delta_L(X) = 
    \begin{cases}
    1, &\text{ if } S_L(\mathbf{X}_n, p_\infty) - S_L(\mathbf{X}_n, p_1)  > T \\
    0 &\quad \text{ otherwise}.
        \end{cases}
\end{equation}
Here,
$$
S_L(\mathbf{X}_n, p) = -\log \prod_{i=1}^n p({X}_i).
$$
is the log score of the sample $(X_1, \dots, X_n)$. The choice of the threshold $T$ depends on the problem formulation. In the variational setting of Neymann and Pearson, the threshold is chosen to set the type-I error. The performance and optimality properties of this test can be found in most standard texts; see, for example,  \cite{moulin2018statistical, lehmann1986testing}. When the densities $p_\infty$ and $p_1$ are not known, the LRT statistic is replaced by a suitable statistic depending on the assumptions on the data. Such families of tests include the generalized likelihood ratio (GLR) test, mixture tests, and nonparametric tests \cite{lehmann1986testing, wasserman2006all}.

\subsection{Limitations of the LRT for Score-Based Models}
In modern machine learning applications, two new classes of models have emerged:
\begin{enumerate}
\item \textit{Unnormalized statistical models}: In these models, we know the densities $p_\infty$ and $p_1$ within normalizing constants. Specifically, we have 
\begin{equation}
\begin{split}
    p_\infty(x) = \frac{\tilde{p}_\infty(x)}{Z_\infty},  \quad \text{ and } \quad 
    p_1(x) = \frac{\tilde{p}_1(x)}{Z_1}.
\end{split}
\end{equation}
Here 
\begin{equation}
\begin{split}
    Z_\infty = \int_x \tilde{p}_\infty(x) dx,  \quad \text{ and  } \quad 
    Z_1 = \int_x \tilde{p}_1(x) dx. 
\end{split}
\end{equation}
are normalizing constants that are hard (or even impossible) to calculate by numerical integration. The unnormalized models $\tilde{p}_\infty(x)$ and $\tilde{p}_1(x)$ are known in precise functional forms. Examples include continuous-valued Markov random fields or undirected graphical models which are used for image modeling  \cite{hyvarinen2005estimation, wuetal-aistat-2023}.  In this case, the LRT can still be implemented. This is because the LRT can be written as 
\begin{equation}
\begin{split}
    \delta_L(X) &= 
    \begin{cases}
    1, &\text{ if } \prod_{i=1}^n\frac{ p_1({X}_i)}{p_\infty({X}_i)}  > T \\
    0 &\quad \text{ otherwise},
        \end{cases}\\
         &=  
    \begin{cases}
    1, &\text{ if } \prod_{i=1}^n\frac{ \tilde{p}_1({X}_i)}{\tilde{p}_\infty({X}_i)}  > T_1 \\
    0 &\quad \text{ otherwise}.
        \end{cases}
        \end{split}
\end{equation}
Here $T_1 = \left(\frac{Z_1}{Z_\infty}\right)^n T$. 
If $T_1$ is chosen to satisfy a constraint on the type-I error (which can be done by sampling from $p_\infty$ using its unnormalized version $\tilde{p}_\infty$), then the LRT test can still be implemented.  

 \item \textit{Score-based models}: In many modern machine learning applications, even $\tilde{p}_\infty(x)$ and $\tilde{p}_1(x)$ are unknown. But, we may learn the scores,
	$$
	\nabla_x \log p_\infty(x) \quad \text{ and 
 }\quad \nabla_x \log p_1(x),
	$$
 from data. This is possible using the idea of \textit{score-matching}. Specifically, these scores can be learned using a deep neural network \cite{hyvarinen2005estimation, song2020score, wuetal-aistat-2023, song2019generative, vincent2011connection, wuetal-UAI-2023}. We note that a score-based model is also unnormalized where the exact form of the unnormalized function is hard to estimate. In this case, the LRT cannot be implemented and a fresh approach is needed. 
\end{enumerate}

\subsection{Score-Based Hypothesis Testing}
A score-based approach to hypothesis testing was taken in \cite{wu2022score}. This test was based on the concept of Hyv\"arinen Score \cite{hyvarinen2005estimation}. To define this score and the corresponding test,
let $p$ and $q$ be two probability densities defined on $\mathcal{X}$. 
The Fisher divergence between $p$ and $q$ is defined as
\begin{align}
    \label{eq:fisher_div}
    \mathbb{D}_{\texttt{F}} (p \; || \; q) \de \mathbb{E}_{{X}\sim p} \left[\frac{1}{2}\left \| \nabla_{x} \log p({X}) - \nabla_{{x}} \log q({X}) \right \|_2^2\right],
\end{align}
whenever the integral is well defined. Under some mild regularity conditions on $p$ and $q$~\cite{hyvarinen2005estimation}, the Fisher divergence can be rewritten as
\begin{align}
    \label{eq:fisher_expansion}
    \mathbb{D}_{\texttt{F}} (p \; || \; q) = \mathbb{E}_{{X}\sim p} \left[\frac{1}{2}\left \| \nabla_{{x}} \log p({X}) \right \|_2^2 + S_{\texttt{H}}( {X}, q)\right],
\end{align}
with 
\begin{align}
    \label{eq:hyv_score}
    S_{\texttt{H}}({X}, q) \de \frac{1}{2} \left \| \nabla_{{x}} \log q({X}) \right \|_2^2 + \Delta_{{x}} \log q({X}),
\end{align}
where $\Delta_{{x}} = \sum_{i=1}^d \frac{\partial^2}{\partial x_i^2}$ denotes the Laplacian operator with respect to ${x} = (x_1, \cdots, x_d)^{\top}$.  The score $S_{\texttt{H}}({X}, q)$ is the Hyv\"arinen score \cite{hyvarinen2005estimation} and is a counterpart of the logarithmic score function $S_L(X,p) = - \log q({X})$ that corresponds to the widely used negative log-likelihood. We further use the notation ${S}_{\texttt{H}}(\mathbf{X}_n, p)$ to denote
\begin{equation}
    S_{\texttt{H}}(\mathbf{X}_n, p) \de \frac{1}{n} \sum_{i=1}^n S_{\texttt{H}}({X}_i, p).
\end{equation}
The score-based test for binary hypothesis testing proposed in \cite{wu2022score} is given by
\begin{equation}
\label{eq:ScoreHT}
    \delta(\mathbf{X}_n) = 
    \begin{cases}
    1, &\text{ if } S_\texttt{H}(\mathbf{X}_n, p_\infty) - S_\texttt{H}(\mathbf{X}_n, p_1)  > T \\
    0 &\quad \text{ otherwise}.
        \end{cases}
\end{equation}
In this paper, we analyze the type-I and type-II error probabilities of this test as $n \to \infty$ using large deviation theory \cite{Cramer, dembo2009large}.  In the test of the paper, we drop the subscript $H$ from the $S_\texttt{H}(\mathbf{X}_n, p)$ and simply refer to the Hyv\"arinen score by 
$S(\mathbf{X}, p)$ for a collection of $n$ data points and by $S({X}, p)$ for a single data point.

Recently, there have been some attempts to perform out-of-distribution (OOD) detection based on score-matching estimates ~\cite{mahmood2020multiscale, kulinski2020feature}.  Note that in OOD detection, the in-distribution samples are assumed to follow a distribution $p(\cdot)$, while the OOD samples are distributed according to another distribution $q(\cdot)$. For an observation $x$,  hypothesis  $\mathcal{H}_0: x \sim p$ is then tested against $\mathcal{H}_1: x \not \sim p$. 
In~\cite{mahmood2020multiscale}, the authors utilized the norm of the gradient of logarithmic likelihood at multiple noise scales for anomaly detection. The authors of~\cite{kulinski2020feature} estimated the Fisher divergence between the null and alternative distributions for feature shift detection assuming that the data is generated according to the null distribution. In \cite{wu2022score}, the scores of null distribution and alternative distributions were used to propose a score-based hypothesis test. As we will be analyzing the performance of this method, we will review its required background and subsequently discuss it below.

\section{Fixed-Sample Error Analysis of Score-Based Test for Binary Hypothesis Testing}
\label{sec:FixedSampleError}
Consider the single-sample score-based simple hypotheses test between densities $p_\infty$ and $p_1$ using $n$ data points $\mathbf{X}_n=(X_1, \dots, X_n)$ (after supressing the subscript $H$ in \eqref{eq:ScoreHT}):
\begin{equation}
    \delta_n \de \delta(\mathbf{X}_n) = 
    \begin{cases}
    1, &\text{ if } S(\mathbf{X}_n, p_\infty) - S(\mathbf{X}_n, p_1)  > T \\
    0 &\quad \text{ otherwise},
        \end{cases}
\end{equation}
where recall that $S(\mathbf{X}_n, p) = \frac{1}{n} \sum_{i=1}^n S({X}_i, p)$.
Thus, choosing $\delta_n=1$ is interpreted as the selection of the alternative hypothesis $\mathcal{H}_1: p=p_1$. 
The type I error (probability of false alarm) is then given by
\begin{equation}
    \alpha_n(\delta_n) \de \mathsf{P}_\infty(\delta_n = 1) = \mathsf{P}_\infty\left(S(\mathbf{X}_n, p_\infty) - S(\mathbf{X}_n, p_1) > T\right).
\end{equation}
This probability can be estimated using Chernoff's bound: for $\theta > 0$
\begin{equation}
    \begin{split}
        \alpha_n(\delta_n) &=\mathsf{P}_\infty\left(S(\mathbf{X}_n, p_\infty) - S(\mathbf{X}_n, p_1) > T\right) \\
        & = \mathsf{P}_\infty \left(\frac{1}{n}
   \sum_{i=1}^n (S(X_i, p_\infty) - S(X_i, p_1)    >  T\right) \\
   & = \mathsf{P}_\infty \left(
   \sum_{i=1}^n (S(X_i, p_\infty) - S(X_i, p_1)    >  n T\right) \\
   &=\mathsf{P}_\infty\left(e^{\theta (\sum_{i=1}^n (S(X_i, p_\infty) - S(X_i, p_1) )  )} > e^{n\theta T}\right) \\
        &\leq e^{-n\theta T} \mathsf{E}_\infty \left [e^{\theta (\sum_{i=1}^n (S(X_i, p_\infty) - S(X_i, p_1) )  )} \right]\\
        &= \left[e^{-\theta T} e^{\log \mathsf{E}_\infty\left[e^{\theta(S(X, p_\infty) - S(X, p_1))}\right]} \right]^n\\
        &=\exp  \left[n \left(\log \mathsf{E}_\infty\left[e^{\theta(S(X, p_\infty) - S(X, p_1))}\right] - \theta T\right)\right],
    \end{split}
\end{equation}
where $\mathsf{E}_\infty(\cdot)$ denotes expectation with respect to $\mathsf{P}_\infty$.

Since this inequality is true for every $\theta > 0$, we can take an infimum over $\theta$ on the right-hand side. 
\begin{equation}
    \begin{split}
        \alpha_n&(\delta_n)\\
        &\leq \inf_{\theta \geq 0}\exp \left(n\log \mathsf{E}_\infty\left[e^{\theta(S(X, p_\infty) - S(X, p_1))}\right] - \theta T\right)\\
        &=\exp \left(n\inf_{\theta \geq 0} \left(\log \mathsf{E}_\infty\left[e^{\theta(S(X, p_\infty) - S(X, p_1))}\right] - \theta T\right)\right)\\
        &=\exp \left(- n\sup_{\theta \geq 0} \left(\theta T - \log \mathsf{E}_\infty\left[e^{\theta(S(X, p_\infty) - S(X, p_1))}\right] \right)\right).
    \end{split}
\end{equation}
Define
$$
\phi(\theta) =  \log \mathsf{E}_\infty\left[e^{\theta(S(X, P_\infty) - S(X, P_1))}\right].
$$
It is well-known that $\phi(\theta)$ is a convex function. Define the Legendre transformation of the convex function $\phi(\theta)$ as
$$
\phi^*(T) = \sup_{\theta \geq 0} \left[ \theta T -\phi(\theta)\right].
$$
Thus, 
\begin{equation}
    \alpha_n (\delta_n) \leq e^{-n\phi^*(T)},
\end{equation}
for all $n \ge 1$. Note that $\phi^*(T)$ is strictly positive if $T$ is larger than the the slope of $\phi(\theta)$ at $\theta=0$, which is
\begin{equation*}
\begin{split}
    \frac{d}{d \theta} \phi(\theta) \bigg|_{\theta=0} &=  \frac{d}{d \theta} \log \mathsf{E}_\infty\left[e^{\theta(S(X, p_\infty) - S(X, p_1))}\right] \bigg|_{\theta=0} \\
    &= - \mathbb{D}_{\texttt{F}}(p_\infty \; \| \; p_1) < 0.
    \end{split}
\end{equation*}
Thus, for $T > - \mathbb{D}_{\texttt{F}}(p_\infty \; \| \; p_1) $,
\begin{equation}
    \alpha_n (\delta_n) \leq e^{-n\phi^*(T)} < 1
\end{equation}
for all $n \ge 1$. Thus,
\begin{equation}
\label{eq:finiteLogError}
\frac{\log \alpha_n(\delta_n)}{n} \leq - \phi^*(T).
\end{equation}
In the next section, we show that this bound $- \phi^*(T)$ is asymptotically tight using large deviation theory.

We note that a similar argument can be made for the type II error by changing the roles of $P_1$ and $P_\infty$. Specifically, we have
\begin{equation}
    \begin{split}
        \beta_n&(\delta_n) =\mathsf{P}_1 \left( S(\mathbf{X}_n, p_1) - S(\mathbf{X}_n, p_\infty) > -T\right) \\
        &\leq \exp \left(-n \sup_{\theta \geq 0} \left(-\theta T - \log \mathsf{E}_1 \left[e^{\theta(S(X, p_1) - S(X, p_\infty))}\right] \right)\right),
    \end{split}
\end{equation}
for all $n \ge 1$, where $\mathsf{E}_1(\cdot)$ denotes expectation with respect to $\mathsf{P}_1$. We conclude that
$$ 
\frac{\log \beta_n(\delta_n)}{n} \leq - \sup_{\theta \geq 0} \left(-\theta T - \log \mathsf{E}_1 \left[e^{\theta(S(X, P_1) - S(X, P_\infty) }\right] \right).
$$

\section{Large Deviation Analysis of the Score-Based Test}
\label{sec:LargeDeviation}

By Cramer's theorem \cite{Cramer}, the type-I error has a positive error exponent given by
\begin{equation*}
\begin{split}
   \frac{1}{n} &\log \mathsf{P}_\infty(\delta_n = 1) \\
   &=  \frac{1}{n} \log \mathsf{P}_\infty \left( \frac{1}{n} \sum_{i=1}^n (S(X_i, p_\infty) - S(X_i, p_1))  > T\right) \\
   &\to \inf_{\theta \geq 0} \left(\log \mathsf{E}_\infty\left[ e^{\theta(S(X_1, p_\infty) - S(X_1, p_1))}\right]  - \theta T\right) \\
   &= -  \sup_{\theta \geq 0} \left(\theta T - \log \mathsf{E}_\infty\left[ e^{\theta(S(X_1, p_\infty) - S(X_1, p_1))}\right]  \right) \\
   &= -\phi^*(T).
   \end{split}
\end{equation*}
Thus, the exponent we calculated in \eqref{eq:finiteLogError} for a sample of size $n$ (using the Chernoff bound)  turns out to be tight in the asymptotic regime as $n \rightarrow \infty$. 

Similarly, by Cramer's theorem again, the type-II error has a positive error exponent given by
\begin{equation*}
\begin{split}
   \frac{1}{n} &\log \mathsf{P}_1(\delta_n = 0) \\&=  \frac{1}{n} \log \mathsf{P}_1 \left( \frac{1}{n} \sum_{i=1}^n (S(X_i, p_\infty) - S(X_i, p_1))  < T\right) \\
   &=  \frac{1}{n} \log \mathsf{P}_1 \left( \frac{1}{n} \sum_{i=1}^n (S(X_i, p_1) - S(X_i, p_\infty))  > -T\right) \\
   &\to \inf_{\theta \geq 0} \left(\log \mathsf{E}_1\left[ e^{\theta(S(X_1, p_1) - S(X_1, p_\infty))}\right]  + \theta T\right) \\
   &= -  \sup_{\theta \geq 0} \left(-\theta T - \log \mathsf{E}_1\left[ e^{\theta(S(X_1, p_1) - S(X_1, p_\infty))}\right]  \right).
   \end{split}
\end{equation*}
Let
$$
- \mathbb{D}_{\texttt{F}}(p_\infty \; \| \; p_1) < T < \mathbb{D}_{\texttt{F}}(p_1\; \| \; p_\infty ).
$$
Then, 
$$
-T > - \mathbb{D}_{\texttt{F}}(p_1 \; \| \; p_\infty),
$$
and this error exponent is also positive. 

\section{Estimating the Error Exponent}
\label{sec:EstimatingErrExpo}
The error exponent $\phi^*(T)$ can be estimated empirically by using numerical simulation to design the hypothesis test. Since we have access to the score $\nabla \log p_\infty(x)$, we can sample from it using Markov Chain Monte Carlo (MCMC) techniques such as Metropolis Adjusted Langevin Algorithm (MALA), Hamilitonian Monte Carlo (HMC), etc. Let $X_1, X_2, \dots, X_m$ be $m$ samples generated using such an MCMC Algorithm. Then the function of $\theta$ given by
$$
\theta T - \log \mathsf{E}_\infty\left[ e^{\theta(S(X_1, p_\infty) - S(X_1, p_1))}\right] 
$$
can be empirically estimated by
$$
\theta T - \log \frac{1}{m} \sum_{k=1}^m \left[ e^{\theta(S(X_k, p_\infty) - S(X_k, p_1))}\right].
$$
The above expression can be optimized over $\theta$ using standard gradient descent algorithms in order to estimate $\phi^*(T)$. The error exponent for type II error can be numerically calculated in an analogous manner.

\vspace{1cm}


\section{Explicit Calculation For Specific Cases}
\label{sec:ExampleCalculations}
We note that for some specific distributions, the above error exponents can be calculated in closed form.
As an example, we consider the standard simple Gaussian hypothesis testing with where the null and alternative hypothesis have the same co-variance matrices. Without loss of generality, we can assume that $P_\infty$ is multidimensional Gaussian $N(0, \Sigma)$ with mean zero and co-variance matrix $\Sigma$ and $P_1$ is $N(\mu, \Sigma)$. Then by direct calculation:
$$ S(X, p_\infty) - S(X, p_1) =  - \frac{\mu^T}{2} \Sigma^{-2} (X-\frac{\mu}{2}).$$
Another straightforward calculation gives
\begin{equation}
    \begin{split}
        \mathsf{E}_\infty\left[e^{\theta(S(X, p_\infty) - S(X, p_1))}\right] &=
\mathsf{E}_\infty\left[e^{-\theta \frac{\mu^T}{2} \Sigma^{-2} (X-\frac{\mu}{2} ) }\right] \\&= e^{ \theta \frac{\mu^T \Sigma^{-2} \mu}{4}} 
e^{\theta^2 \frac{\mu^T\Sigma^{-3} \mu}{8}}.
    \end{split}
\end{equation}
We thus have
\begin{equation}
    \begin{split}
\theta T &- \log \mathsf{E}_\infty\left[ e^{\theta(S(X, p_\infty) - S(X, p_1))}\right] \\
&= \theta T  - \theta \frac{\mu^T \Sigma^{-2} \mu}{4} - \theta^2 \frac{\mu^T\Sigma^{-3} \mu}{8}.
   \end{split}
\end{equation}
If $T < \frac{\mu^T \Sigma^{-2} \mu}{4}$ then the maxima is achieved at $\theta = 0$ and the type I error exponent is given by $\phi^*(T) = 0$. Otherwise, the maxima is achieved at 
$\theta = \frac{4T - \mu^T \Sigma^{-2} \mu} {\mu^T\Sigma^{-3} \mu}$ and
the type I error exponent is given by
$$\phi^*(T) =   \frac{ (4T - \mu^T \Sigma^{-2} \mu )^2 } {8 \mu^T\Sigma^{-3} \mu}, $$ 
which is quadratic in $T$, which is the same behavior as the log-likelihood ratio test. 

\section{Numerical Experiments}
\label{sec:NumericalResults}
In this section, numerical experiments are conducted using various datasets in order to numerically demonstrate the effectiveness of our large deviation analysis of score-based hypothesis testing.
 Our experiments utilize both synthetic and real-world data. 
 \begin{enumerate}
     \item \textit{Synethetic datasets}:  For synthetic datasets, we consider samples generated from multivariate normal distributions, exponential family, and Gauss-Bernoulli Restricted Boltzmann Machines (RBMs)~\cite{wu2022score}. When the exact likelihood is available, as in the case of multivariate normal distributions and the exponential family, we compare the performance of the Hyvarinen Score Test (HST) with that of the Likelihood Ratio Test (LRT). For each distribution type, we perform $10$ runs, with each run producing a datasets of size $10,000$ created by sampling from both the null and alternative distributions.
     \item \textit{Real datasets}: For real-world datasets, we utilize the KDD Cup’99 dataset~\cite{lippmann2000analysis}. This dataset contains various types of simulated adversarial network attacks and is typically used for bench-marking network intrusion detection algorithms. 
 \end{enumerate}
 
 For all experiments, we evaluate the \textit{Positive Error Exponent} (for Type I error) and \textit{Negative Error Exponent} (for Type II error) for various thresholds. Using numerical calculations, we compare the estimates of the theoretical results obtained from the large deviation analysis given above, with the empirical error exponent obtained from hypothesis tests of size $n$ by sampling with replacement (ranging from $n=1$ to $n=128$). The threshold for calculating the theoretical limit is obtained empirically with $n=1$.

\subsection{Details of Experimental Setup}
We now discuss the experiments in detail.  

\paragraph{Multivariate normal distribution}
For the multivariate normal distribution data, we consider the bivariate normal distribution $\mathcal{N}(\mu, \Sigma)$, performing tests on $\mu$ with known $\Sigma$, and vice versa. The null hypothesis $P_{\infty}$ is: The mean $\mu_0$ (respectively the covariance $\Sigma_0$) is equal to its true value $\mu^{\star}$ (or $\Sigma^{\star}$). The corresponding values of the alternative hypothesis $P_{1}$ are assigned by adding i.i.d. samples of a noise term $\mathcal{N}(0,\sigma_{ptb}^2)$ to coordinates of $\mu_{\infty}$ (or by adding a i.i.d. samples of a log-Normal noise to diagonal elements of $\Sigma_{\infty}$, where the logarithm of each noise term is distributed according to a Normal distribution $\mathcal{N}(0,\sigma_{ptb}^2)$). In our simulations, we let $\mu_{\infty}=(0,0)^{T}$ (respectively $\Sigma_{\infty}=\left[ {\begin{array}{cc}
   1 & 0.7 \\
   0.7 & 1\\
  \end{array} } \right]$) for the null hypotheses in each trial, testing against alternatives $\mu_1$ (respectively $\Sigma_1$) as described above with $\sigma_{ptb}=0.01$.

\paragraph{Exponential family} For the exponential family dataset, we use the random variable $\mathbf{x}\in \mathbb{R}^d$ distributed according to the un-normalized pdf 
\begin{align}
    p_{\tau}(\mathbf{x}) \propto \exp\left\{-\tau\left(\sum_{i=1}^dx_i^4+\sum_{1\leq i\leq d, i\leq j\leq d}x_i^2x_j^2\right)\right\},
\end{align}
where $\tau\in \mathcal{T}\subset\mathbb{R}^{+}$ is the model parameter. This subfamily of the exponential family correspond to pairwise interaction graphical models~\cite{yu2016statistical}. We first consider the hypothesis test $\mathcal{H}_{\infty}:\tau = \tau_{\infty}=1$ versus $\mathcal{H}_1:\tau = \tau_1=\tau_0+\tau_{ptb}$, with $\tau_{ptb}=0.01$. Hamiltonian Monte Carlo (HMC) is used to generate samples from the un-normalized density functions. In order to perform LRT, we compute the normalizing constant by numerical integration.

\paragraph{RBM} We also consider the RBM~\cite{LeCun2006ATO} model which is defined based on an undirected bi-partite graphical model consisting of hidden and visible variables. The Gauss-Bernoulli RBM has binary-valued hidden variables $\mathbf{h}\in \{0,1\}^{d_h}$ and real-valued visible variables $\mathbf{x}\in R^{d_x}$ with joint distribution
\begin{align*}
    p(\mathbf{x}, \mathbf{h}) = &\frac{1}{Z_\theta}\text{exp} \left\{
    -\left(\frac{1}{2}\sum_{i=1}^{d_x}\sum_{j=1}^{d_h}\frac{x_i}{\sigma_i}W_{ij}h_j \right. \right.\\
    &\left. \left. \quad \quad \quad \quad +\sum_{i=1}^{d_x}b_ix_i+\sum_{j=1}^{d_h}c_jh_j-\frac{1}{2}\sum_{i=1}^{d_{x}} \frac{x_{i}^{2}}{\sigma_{i}^{2}}\right)
    \right\},
\end{align*}
where  $\theta = (\mathbf{W}, \mathbf{b}, \mathbf{c})$ are model parameters and $Z_\theta$ is the normalizing constant. We set $\sigma_i=1$ for all $i=1,\dots, d_x$ in our experiments. The probability of the visible variable $\mathbf{x}$ written as $p(\mathbf{x})= \sum_{h\in \{0,1\}^{d_h}}p(\mathbf{x}, \mathbf{h}) = \frac{1}{Z_\theta}\exp\{-F_{\theta}(\mathbf{x})\}$, where $F_{\theta}(\mathbf{x})$ is the free energy given by 
\begin{align}
    F_{\theta}(\mathbf{x}) =& \frac{1}{2}\sum_{i=1}^{d_x} (x_{i}-b_i)^{2}-\sum_{j=1}^{d_h} \operatorname{Softplus}\left(\sum_{i=1}^{d_x} W_{i j}x_{i}+b_{j}\right).
\end{align}
The $\operatorname{Softplus}$ function is defined as $\operatorname{Softplus}(t) \de \log(1+\exp(t))$ with a default scale parameter $\beta=1$. The corresponding Hyv\"arinen score $S_{\texttt{H}}(\mathbf{X}_n, \theta)$ is given by
\begin{align*}
S_{\texttt{H}}(\mathbf{X}_n, \theta)= \sum_{n=1}^{n} \sum_{i=1}^{d_x} & \left[\frac{1}{2}\left(x_{in}-b_{i}+\sum_{j=1}^{d_h} W_{ij} \delta_{jn}\right)^2 \right. \\
& \quad \quad \left.+\sum_{j=1}^{d_h} W_{i j}^{2} \delta_{jn}\left(1-\delta_{jn}\right)-1\right],
\end{align*}
where $\delta_{jn} \de \operatorname{Sigmoid}(\sum_{i=1}^{d_x} W_{i j}x_{in}+b_{j})$. The $\operatorname{Sigmoid}$ function is defined as $\operatorname{Sigmoid}(t) \de (1+\exp(-t))^{-1}$. 
We randomly draw the weight matrix $\mathbf{W}_0\in \mathbb{R}^{d_x}\times\mathbb{R}^{d_h}$ from the standard Normal distribution. In our experiments, we set the dimension of visible variables to $d_x=50$ and hidden variables to $d_h=40$. The weight matrix of the alternative hypothesis is constructed by perturbing elements of $\mathbf{W}_0$  with i.i.d samples of Normal distribution $\mathcal{N}(0,\sigma_{ptb}^2)$ with $\sigma_{ptb}=0.01$. Samples of RBMs are drawn using Gibbs sampling with $1000$ RBM iterations in order to ensure convergence.

\paragraph{KDD Cup} For the KDD Cup'99 dataset, the `normal' traffic network is treated as the null hypothesis, against various types of adversarial network attacks as alternative hypotheses. A Gauss-Bernoulli RBM is trained with all available training data of the `normal' traffic network to estimate an unnormalized version of the null distribution \( P_{\infty} \). Hypothesis tests are conducted against various alternative hypotheses, including `back', `ipsweep’, `neptune’, and others. The `unknown' adversarial network attack combines all types of attacks with a dataset size of $\le 100$. To provide a comprehensive overview, the number of available data points of the network traffic in the KDD Cup'99 dataset are depicted in Table~\ref{tab:stats}.

\begin{table}[t]
\centering
\caption{Statistics of various network traffic in KDD Cup'99 dataset.}
\label{tab:stats}
\resizebox{\columnwidth}{!}{%
\begin{tabular}{@{}ccccccc@{}}
\toprule
normal & neptune & back & teardrop & satan & warezclient\\ \midrule
87832 & 51820 & 968 & 918 & 906 & 893\\ \bottomrule
\bottomrule
ipsweep & smurf & portsweep & pod & nmap & unknown \\ \midrule
651 & 641 & 416 & 206 & 158 & 177 \\ \bottomrule
\end{tabular}%
}
\end{table}
We conducted two sets of experiments. In the first set, we explicitly use the \textit{un-normalized} null ($P_{\infty}$) and alternative ($P_1$) distributions from true data distributions of synthetic data. The synthetic data experiments include normal distributions, exponential family, and Gauss-Bernoulli RBM. In the second set of simulations (the Gauss-Bernoulli RBM and KDD Cup’99 datasets), we fit the alternative distribution $P_1$ using training data of size $N$, with cold-start from the null distribution. For the null distribution, we use the explicit un-normalized data distribution for the synthetic Gauss-Bernoulli RBM experiment. For the KDD Cup’99 dataset experiment, we fit the unnormalized null distribution with a Gauss-Bernoulli RBM using all the available training data of the `normal' traffic network.  In order to demonstrate the robustness of our results to the sampling of alternative training data, we produced $100$ alternative distributions, each trained with $N$ alternative data samples. For each test, one of the alternative distributions is randomly selected. Furthermore, we conducted ablation studies on the training data size $N$ in order to demonstrate the influence of the underfitting/overfitting of the alternative distribution on our large deviation analysis results. This is important because in many practical applications (such as detecting adversarial network attacks), alternative data are much more scarce than the null data.

\subsection{Experimental Results}
\label{sec:experiments}
\begin{figure*}[htbp]
    \centering
    \includegraphics[width=1\linewidth]{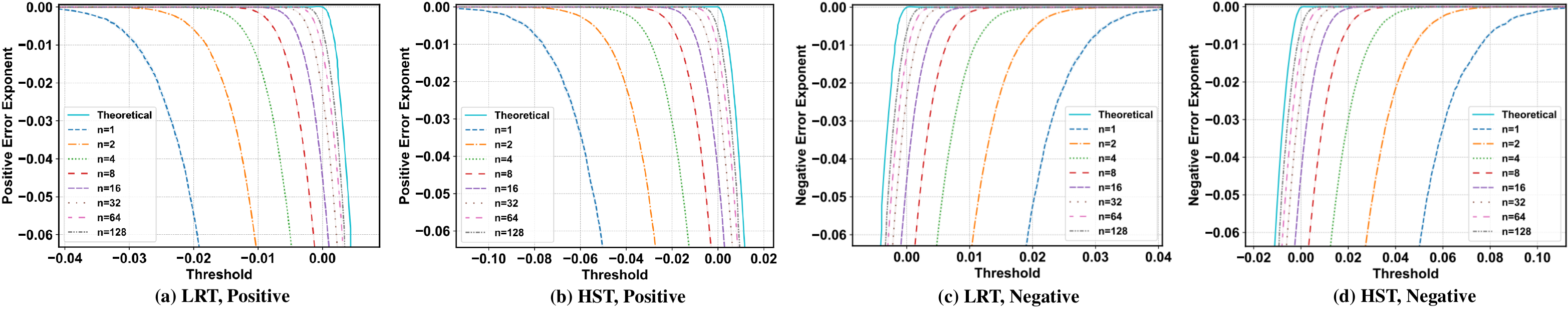}
    \vspace{-0.4cm}
    \caption{Large deviation analysis of likelihood-based and sore-based hypothesis testing for multivariate normal distribution with perturbation on $\mu$ and $\sigma_{ptb} = 0.01$.}
    \label{fig:ptb_mvn_0.01-0.0}
\end{figure*}

\begin{figure*}[htbp]
    \centering
    \includegraphics[width=1\linewidth]{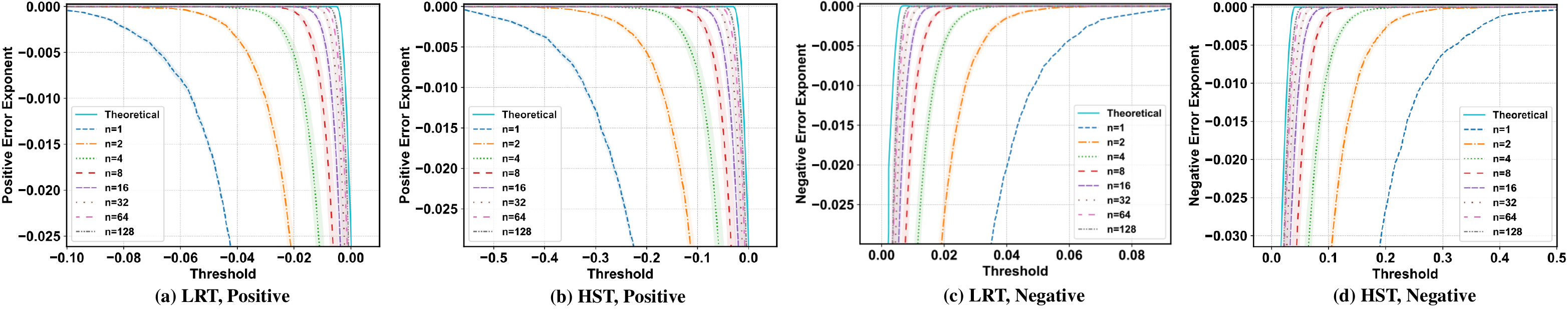}
    \vspace{-0.4cm}
    \caption{Large deviation analysis of likelihood-based and sore-based hypothesis testing for multivariate normal distribution with perturbation on $\sigma$ and $\sigma_{ptb} = 0.01$.}
    \label{fig:ptb_mvn_0.0-0.01}
\end{figure*}

In our first set of experiments (as described above), we evaluate the performance of the Hyvarinen Score Test (HST) and Likelihood Ratio Test (LRT) for explicit data distributions. These distributions include multivariate normal distributions, the exponential family, and Gauss-Bernoulli Restricted Boltzmann Machines (RBMs). Results from these distributions are presented in Figures~\ref{fig:ptb_mvn_0.01-0.0} to~\ref{fig:ptb_mvn_0.0-0.01}. These results are consistent with our theoretical analysis given in Section~\ref{sec:score_matching_test}. Specifically, as the sample size $n$ in the composite hypothesis tests is increased, both the empirical positive and negative error exponents  converge to their corresponding theoretical limits.

\begin{figure*}[htbp]
    \centering
    \includegraphics[width=1\linewidth]{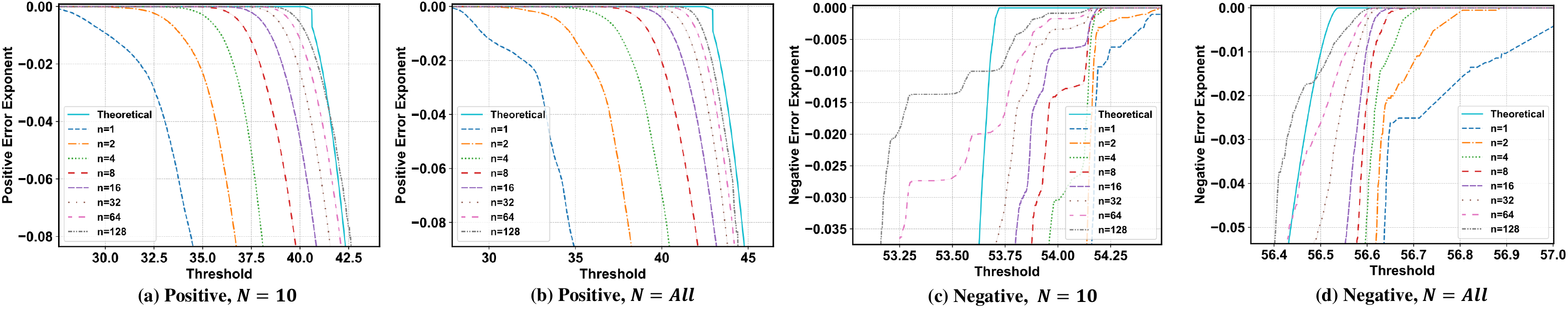}
    \vspace{-0.4cm}
    \caption{Large deviation analysis of sore-based hypothesis testing for `back’ attack on KDD Cup’99 dataset ($P_1$ fitted with Gauss-Bernoulli RBM and $N$ samples).}
    \label{fig:ds_kddcup_back}
\end{figure*}

\begin{figure*}[htbp]
    \centering
    \includegraphics[width=1\linewidth]{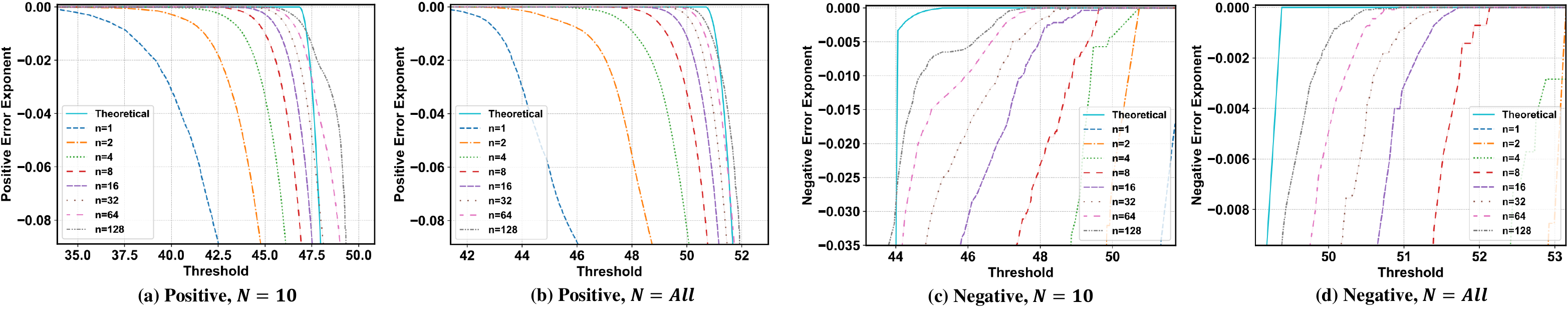}
    \vspace{-0.4cm}
    \caption{Large deviation analysis of sore-based hypothesis testing for `unknown’ attack on KDD Cup’99 dataset ($P_1$ fitted with Gauss-Bernoulli RBM and $N$ samples).}
    \label{fig:ds_kddcup_unknown}
\end{figure*}

In the second set of experiments (as described above), we focus on scenarios where the alternative distribution $P_1$ is fit using the training data. These results are depicted in Figures~\ref{fig:ptb_rbm_0.01} through~\ref{fig:ds_kddcup_unknown}. In these experiments, we observed that the empirical positive and negative error exponents occasionally exceed the theoretical limits as $n$ grows large. However, this trend disappears as the number of samples $N$ (used to fit the alternative distribution) increases. As $N$ increases (for instance from $10$ to $100$ in Figure~\ref{fig:ptb_rbm_0.01} and from $10$ to $All$ in Figure~\ref{fig:ds_kddcup_back}), the empirical results more frequently adhere to the theoretical limits, aligning with our expectation. Furthermore, we also observe that when $N$ increases, the empirical error exponent becomes more convex. These results suggest that a more accurately fitted alternative distribution facilitates the convergence of the empirical error exponents towards the theoretical limits. Further research could investigate the influence of the fitting accuracy of the alternative distribution on the applicability and effectiveness of the large deviation analysis.

\begin{figure*}[htbp]
    \centering
    \includegraphics[width=1\linewidth]{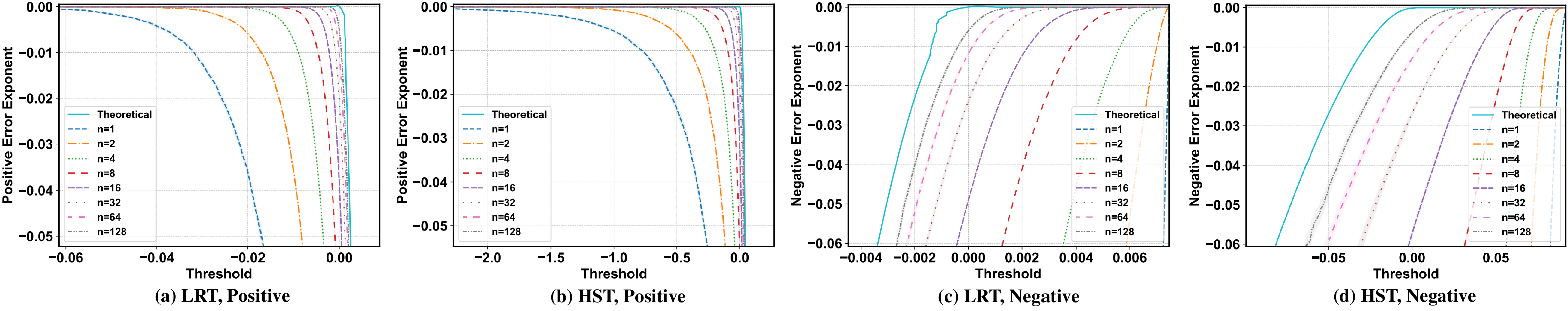}
    \vspace{-0.4cm}
    \caption{Large deviation analysis of likelihood-based and sore-based hypothesis testing for exponential family with perturbation on $\tau$ and $\sigma_{ptb} = 0.01$.}
    \label{fig:ptb_exp_0.01}
\end{figure*}

\begin{figure}[htbp]
    \centering
    \includegraphics[width=1\linewidth]{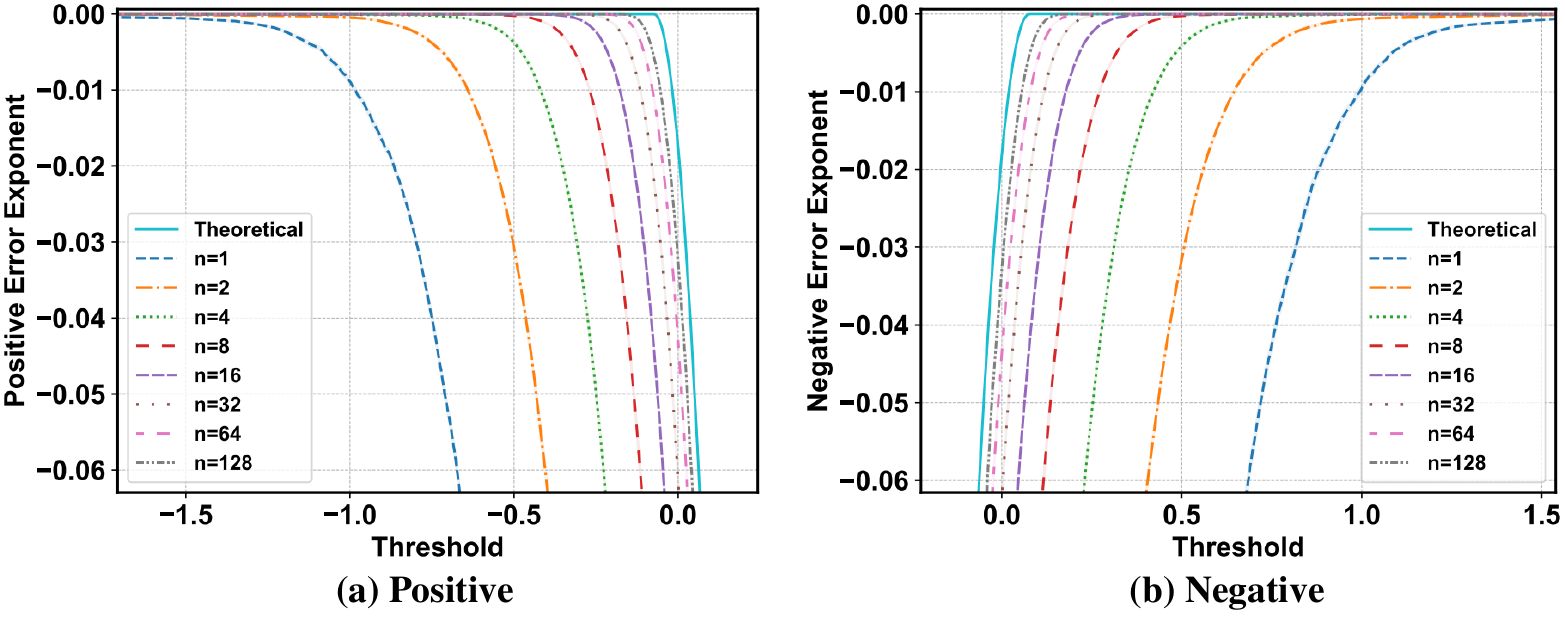}
    \vspace{-0.4cm}
    \caption{Large deviation analysis of sore-based hypothesis testing for Gauss-Bernoulli RBM with perturbation on $W$ and $\sigma_{ptb} = 0.01$.}
    \label{fig:ptb_rbm_0.01}
\end{figure}

\begin{figure*}[htbp]
    \centering
    \includegraphics[width=1\linewidth]{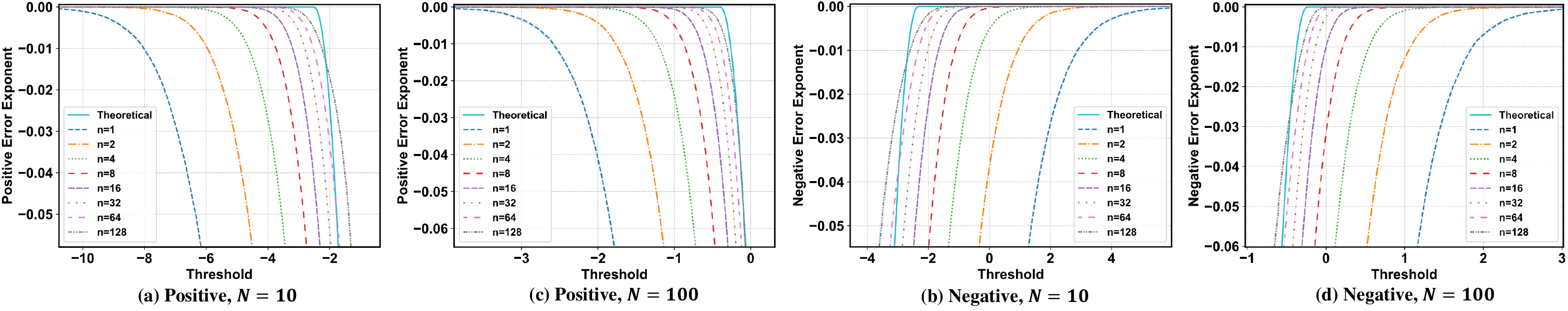}
    \vspace{-0.4cm}
    \caption{Large deviation analysis of sore-based hypothesis testing for Gauss-Bernoulli RBM ($P_1$ fitted with Gauss-Bernoulli RBM and $N$ samples) with perturbation on $W$ and $\sigma_{ptb} = 0.01$.}
    \label{fig:ptb_rbm_0.01}
\end{figure*}

\begin{figure*}[htbp]
    \centering
    \includegraphics[width=1\linewidth]{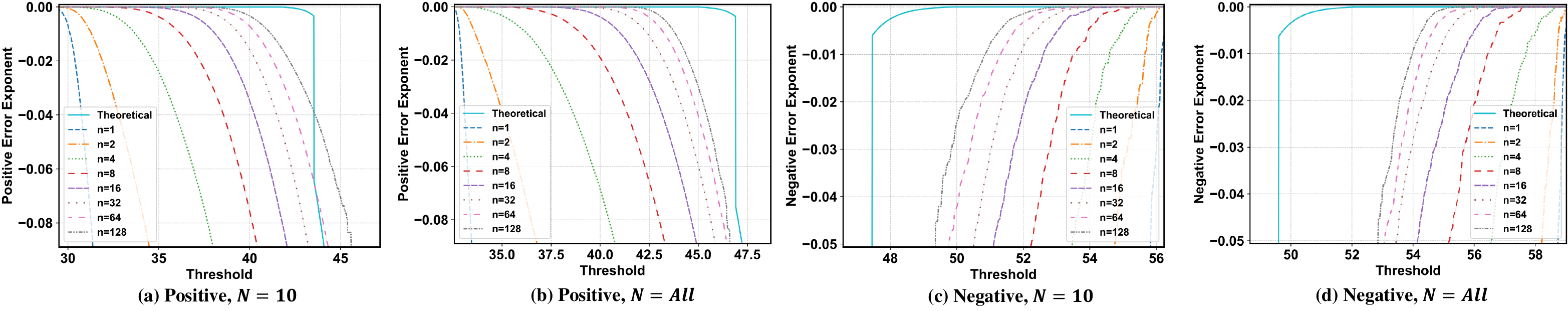}
    \vspace{-0.4cm}
    \caption{Large deviation analysis of sore-based hypothesis testing for `ipsweep’ attack on KDD Cup’99 dataset ($P_1$ fitted with Gauss-Bernoulli RBM and $N$ samples).}
    \label{fig:ds_kddcup_ipsweep}
\end{figure*}

\begin{figure*}[htbp]
    \centering
    \includegraphics[width=1\linewidth]{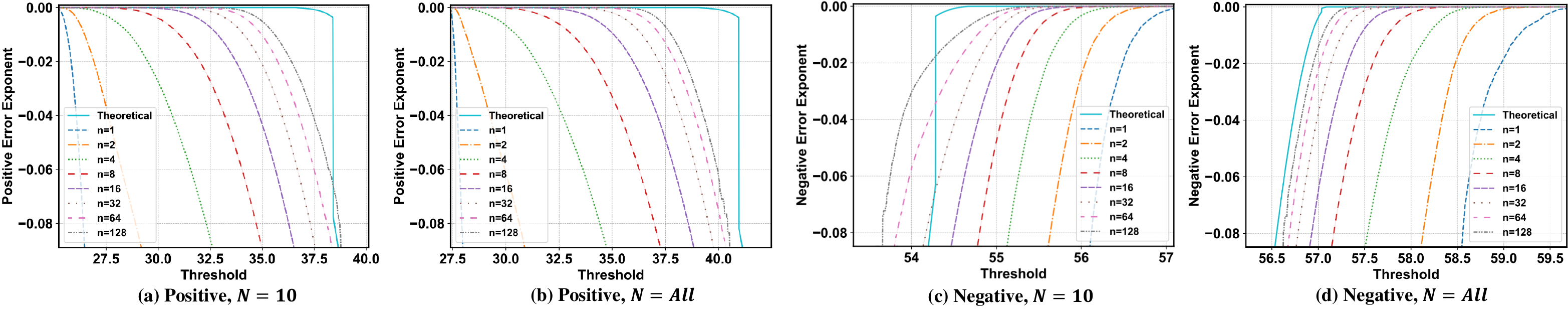}
    \vspace{-0.4cm}
    \caption{Large deviation analysis of sore-based hypothesis testing for `neptune’ attack on KDD Cup’99 dataset ($P_1$ fitted with Gauss-Bernoulli RBM and $N$ samples).}
    \label{fig:ds_kddcup_neptune}
\end{figure*}

\begin{figure*}[htbp]
    \centering
    \includegraphics[width=1\linewidth]{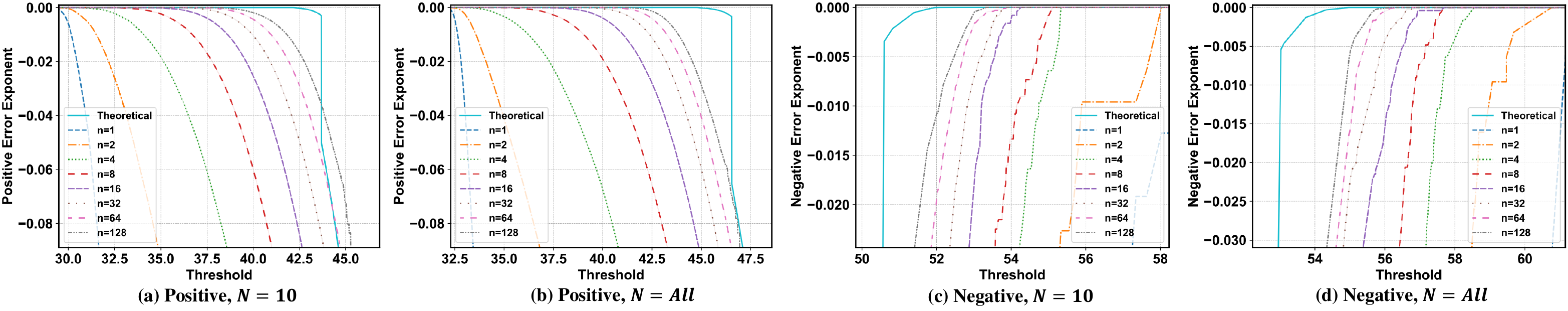}
    \vspace{-0.4cm}
    \caption{Large deviation analysis of sore-based hypothesis testing for `nmap’ attack on KDD Cup’99 dataset ($P_1$ fitted with Gauss-Bernoulli RBM and $N$ samples).}
    \label{fig:ds_kddcup_nmap}
\end{figure*}

\begin{figure*}[htbp]
    \centering
    \includegraphics[width=1\linewidth]{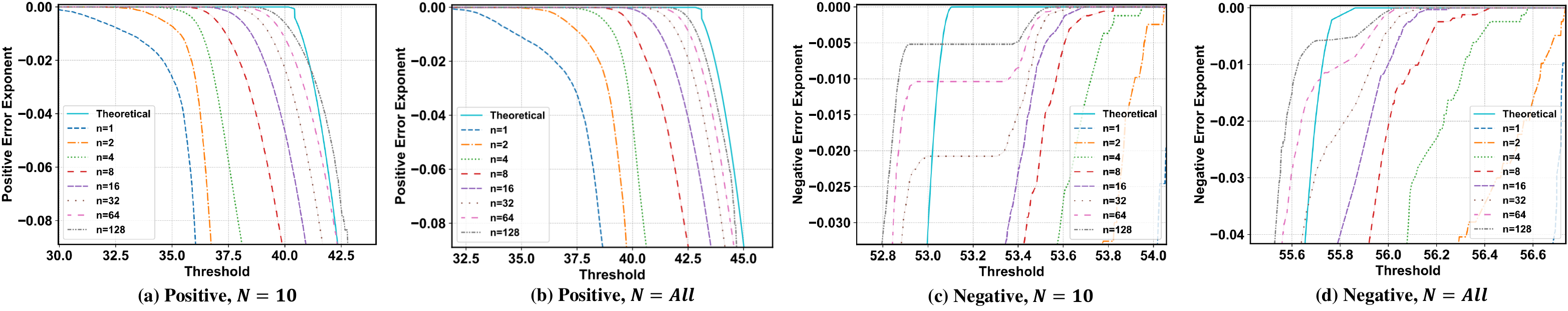}
    \vspace{-0.4cm}
    \caption{Large deviation analysis of sore-based hypothesis testing for `pod’ attack on KDD Cup’99 dataset ($P_1$ fitted with Gauss-Bernoulli RBM and $N$ samples).}
    \label{fig:ds_kddcup_pod}
\end{figure*}

\begin{figure*}[htbp]
    \centering
    \includegraphics[width=1\linewidth]{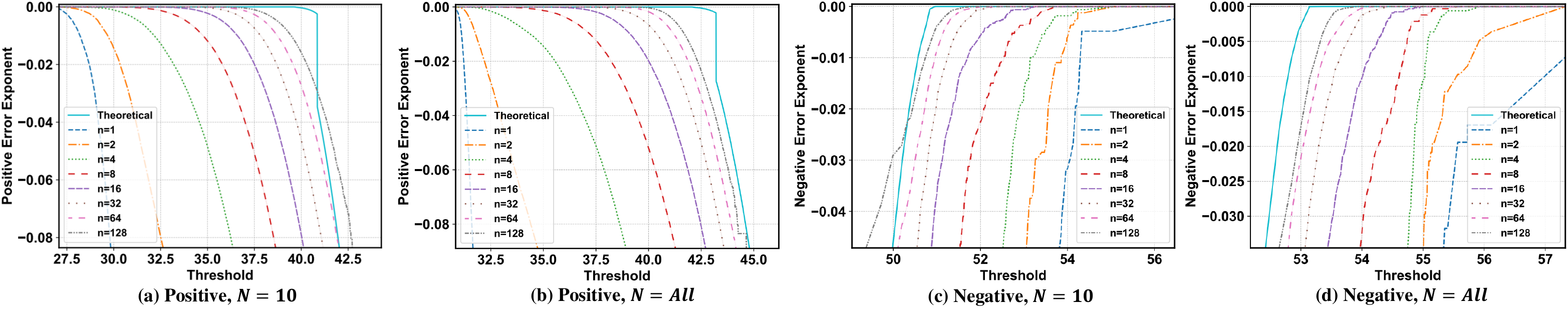}
    \vspace{-0.4cm}
    \caption{Large deviation analysis of sore-based hypothesis testing for `portsweep’ attack on KDD Cup’99 dataset ($P_1$ fitted with Gauss-Bernoulli RBM and $N$ samples).}
    \label{fig:ds_kddcup_portsweep}
\end{figure*}

\begin{figure*}[htbp]
    \centering
    \includegraphics[width=1\linewidth]{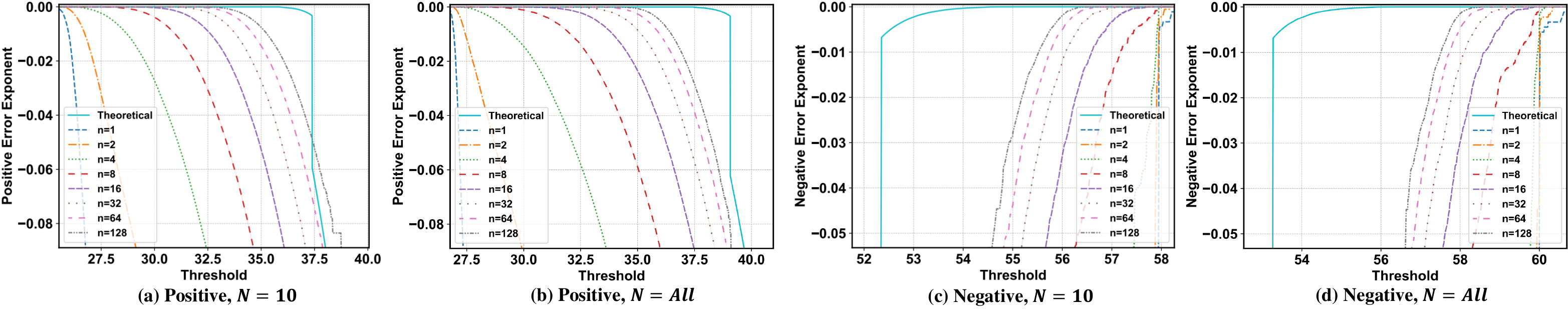}
    \vspace{-0.4cm}
    \caption{Large deviation analysis of sore-based hypothesis testing for `satan’ attack on KDD Cup’99 dataset ($P_1$ fitted with Gauss-Bernoulli RBM and $N$ samples).}
    \label{fig:ds_kddcup_satan}
\end{figure*}

\begin{figure*}[htbp]
    \centering
    \includegraphics[width=1\linewidth]{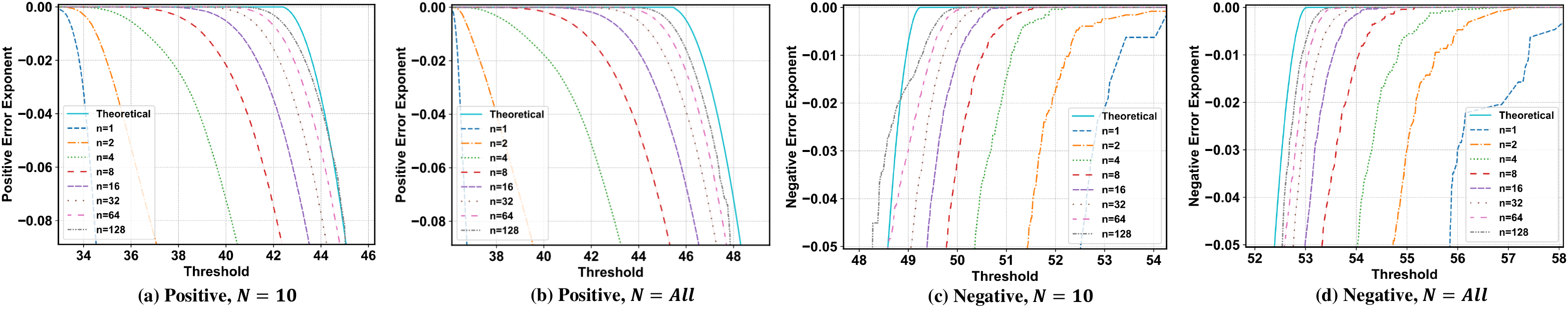}
    \vspace{-0.4cm}
    \caption{Large deviation analysis of sore-based hypothesis testing for `smurf’ attack on KDD Cup’99 dataset ($P_1$ fitted with Gauss-Bernoulli RBM and $N$ samples).}
    \label{fig:ds_kddcup_smurf}
\end{figure*}

\begin{figure*}[htbp]
    \centering
    \includegraphics[width=1\linewidth]{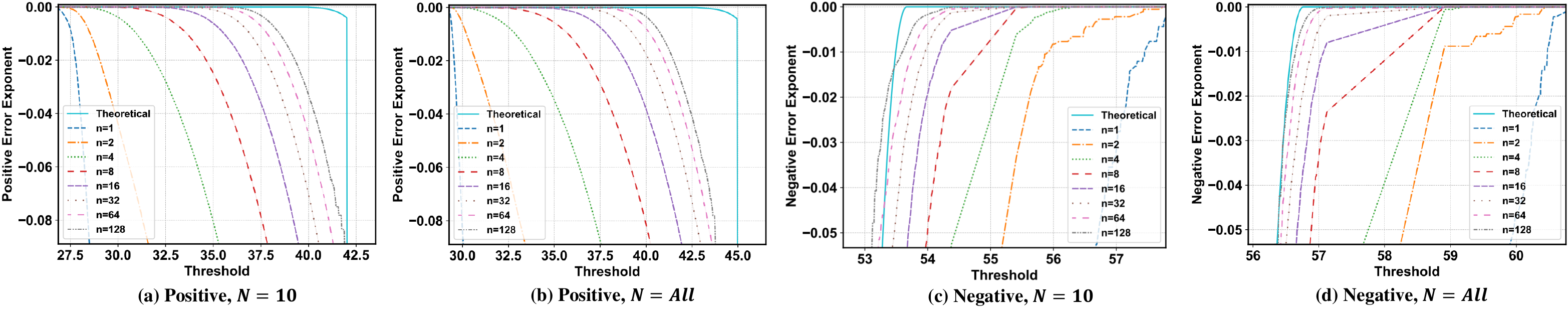}
    \vspace{-0.4cm}
    \caption{Large deviation analysis of sore-based hypothesis testing for `teardrop’ attack on KDD Cup’99 dataset ($P_1$ fitted with Gauss-Bernoulli RBM and $N$ samples).}
    \label{fig:ds_kddcup_teardrop}
\end{figure*}

\begin{figure*}[htbp]
    \centering
    \includegraphics[width=1\linewidth]{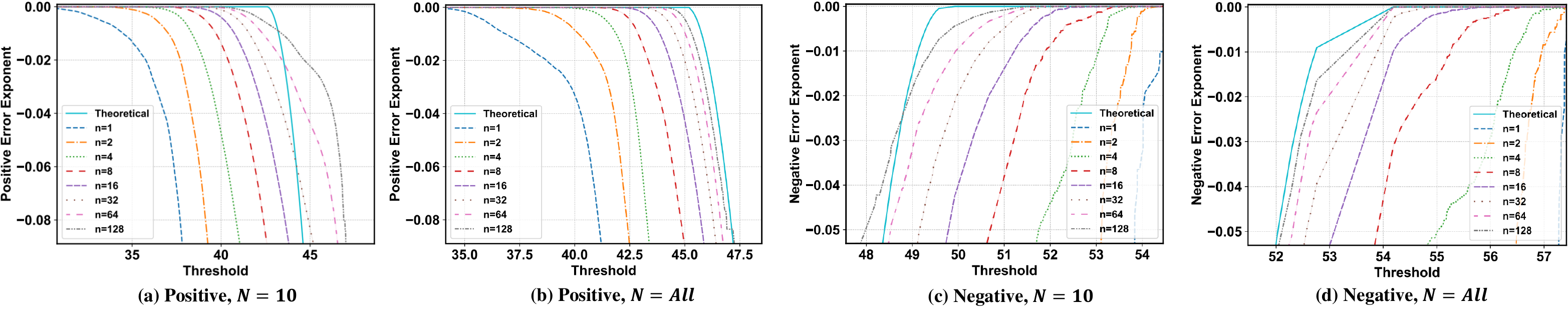}
    \vspace{-0.4cm}
    \caption{Large deviation analysis of sore-based hypothesis testing for `warezclient’ attack on KDD Cup’99 dataset ($P_1$ fitted with Gauss-Bernoulli RBM and $N$ samples).}
    \label{fig:ds_kddcup_warezclient}
\end{figure*}

\section{Conclusion}
\label{sec:conclusion}
In this work, we analyzed the performance of score-based hypothesis testing method \cite{wu2022score}. We derived upper bounds on Type I and II error probabilities, and proved that the exponents of our error bounds become precise in large sample size regimes. We calculated these error exponents numerically for a variety of scenarios of interest. Our experiments using both synthetic and real-world data demonstrated that the empirical error exponents follow our theoretical analysis. Future research may focus on the impact of lack of accuracy in fitting the alternative distribution on the applicability and effectiveness of large deviation analysis.

\balance
\bibliographystyle{IEEEtranN}
\bibliography{reference}

\begin{thebibliography}{21}
\providecommand{\natexlab}[1]{#1}
\providecommand{\url}[1]{#1}
\csname url@samestyle\endcsname
\providecommand{\newblock}{\relax}
\providecommand{\bibinfo}[2]{#2}
\providecommand{\BIBentrySTDinterwordspacing}{\spaceskip=0pt\relax}
\providecommand{\BIBentryALTinterwordstretchfactor}{4}
\providecommand{\BIBentryALTinterwordspacing}{\spaceskip=\fontdimen2\font plus
\BIBentryALTinterwordstretchfactor\fontdimen3\font minus
  \fontdimen4\font\relax}
\providecommand{\BIBforeignlanguage}[2]{{%
\expandafter\ifx\csname l@#1\endcsname\relax
\typeout{** WARNING: IEEEtranN.bst: No hyphenation pattern has been}%
\typeout{** loaded for the language `#1'. Using the pattern for}%
\typeout{** the default language instead.}%
\else
\language=\csname l@#1\endcsname
\fi
#2}}
\providecommand{\BIBdecl}{\relax}
\BIBdecl

\bibitem[Wu et~al.(2022)Wu, Diao, Elkhalil, Ding, and Tarokh]{wu2022score}
S.~Wu, E.~Diao, K.~Elkhalil, J.~Ding, and V.~Tarokh, ``Score-based hypothesis
  testing for unnormalized models,'' \emph{IEEE Access}, vol.~10, pp.
  71\,936--71\,950, 2022.

\bibitem[Hyv{\"a}rinen(2005)]{hyvarinen2005estimation}
A.~Hyv{\"a}rinen, ``Estimation of non-normalized statistical models by score
  matching.'' \emph{J. Mach. Learn. Res.}, vol.~6, no.~4, 2005.

\bibitem[Hyv{\"a}rinen(2007)]{hyvarinen2007some}
------, ``Some extensions of score matching,'' \emph{Comput. Stat. Data Anal.},
  vol.~51, no.~5, pp. 2499--2512, 2007.

\bibitem[Song et~al.(2020)Song, Sohl-Dickstein, Kingma, Kumar, Ermon, and
  Poole]{song2020score}
Y.~Song, J.~Sohl-Dickstein, D.~P. Kingma, A.~Kumar, S.~Ermon, and B.~Poole,
  ``Score-based generative modeling through stochastic differential
  equations,'' \emph{arXiv preprint arXiv:2011.13456}, 2020.

\bibitem[Vahdat et~al.(2021)Vahdat, Kreis, and Kautz]{vahdat2021score}
A.~Vahdat, K.~Kreis, and J.~Kautz, ``Score-based generative modeling in latent
  space,'' \emph{Advances in Neural Information Processing Systems (NeurIPS)},
  vol.~34, pp. 11\,287--11\,302, 2021.

\bibitem[Koller and Friedman(2009)]{graphic_models}
D.~Koller and N.~Friedman, \emph{Probabilistic graphical models: principles and
  techniques}.\hskip 1em plus 0.5em minus 0.4em\relax The MIT Press, 2009.

\bibitem[LeCun et~al.(2006)LeCun, Chopra, Hadsell, Ranzato, and
  Huang]{LeCun2006ATO}
Y.~LeCun, S.~Chopra, R.~Hadsell, M.~Ranzato, and F.~Huang, ``A tutorial on
  energy-based learning,'' in \emph{Predicting structured data}.\hskip 1em plus
  0.5em minus 0.4em\relax The MIT Press, 2006, vol.~1.

\bibitem[Papamakarios et~al.(2021)Papamakarios, Nalisnick, Rezende, Mohamed,
  and Lakshminarayanan]{Papamakarios2021NormalizingFF}
G.~Papamakarios, E.~T. Nalisnick, D.~J. Rezende, S.~Mohamed, and
  B.~Lakshminarayanan, ``Normalizing flows for probabilistic modeling and
  inference,'' \emph{J. Mach. Learn. Res.}, vol.~22, pp. 57:1--57:64, 2021.

\bibitem[Cerf and Petit(2011)]{Cramer}
R.~Cerf and P.~Petit, ``A short proof of cram´er’s theorem in
  $\mathbb{R}$,'' \emph{The American Mathematical Monthly}, vol. 118, no.~10,
  pp. 925--931, 2011.

\bibitem[Dembo and Zeitouni(2009)]{dembo2009large}
A.~Dembo and O.~Zeitouni, \emph{Large deviations techniques and
  applications}.\hskip 1em plus 0.5em minus 0.4em\relax Springer Science \&
  Business Media, 2009, vol.~38.

\bibitem[Moulin and Veeravalli(2018)]{moulin2018statistical}
P.~Moulin and V.~V. Veeravalli, \emph{Statistical Inference for Engineers and
  Data Scientists}.\hskip 1em plus 0.5em minus 0.4em\relax Cambridge University
  Press, 2018.

\bibitem[Lehmann et~al.(1986)Lehmann, Romano, and Casella]{lehmann1986testing}
E.~L. Lehmann, J.~P. Romano, and G.~Casella, \emph{Testing statistical
  hypotheses}.\hskip 1em plus 0.5em minus 0.4em\relax Springer, 1986, vol.~3.

\bibitem[Wasserman(2006)]{wasserman2006all}
L.~Wasserman, \emph{All of nonparametric statistics}.\hskip 1em plus 0.5em
  minus 0.4em\relax Springer Science \& Business Media, 2006.

\bibitem[Wu et~al.(2023{\natexlab{a}})Wu, Diao, Banerjee, Ding, and
  Tarokh]{wuetal-aistat-2023}
S.~Wu, E.~Diao, T.~Banerjee, J.~Ding, and V.~Tarokh, ``Score-based change point
  detection for unnormalized models,'' \emph{International Conference on
  Artificial Intelligence and Statistics (AISTATS)}, 2023.

\bibitem[Song and Ermon(2019)]{song2019generative}
Y.~Song and S.~Ermon, ``Generative modeling by estimating gradients of the data
  distribution,'' \emph{Advances in neural information processing systems},
  vol.~32, 2019.

\bibitem[Vincent(2011)]{vincent2011connection}
P.~Vincent, ``A connection between score matching and denoising autoencoders,''
  \emph{Neural computation}, vol.~23, no.~7, pp. 1661--1674, 2011.

\bibitem[Wu et~al.(2023{\natexlab{b}})Wu, Diao, Banerjee, Ding, and
  Tarokh]{wuetal-UAI-2023}
S.~Wu, E.~Diao, T.~Banerjee, J.~Ding, and V.~Tarokh, ``Robust quickest change
  detection for unnormalized models,'' \emph{Conference on Uncertainty in
  Artificial Intelligence (UAI)}, 2023.

\bibitem[Mahmood et~al.(2020)Mahmood, Oliva, and Styner]{mahmood2020multiscale}
A.~Mahmood, J.~Oliva, and M.~A. Styner, ``Multiscale score matching for
  out-of-distribution detection,'' in \emph{International Conference on
  Learning Representations}, 2020.

\bibitem[Kulinski et~al.(2020)Kulinski, Bagchi, and
  Inouye]{kulinski2020feature}
S.~Kulinski, S.~Bagchi, and D.~I. Inouye, ``Feature shift detection: Localizing
  which features have shifted via conditional distribution tests,''
  \emph{Advances in neural information processing systems}, vol.~33, pp.
  19\,523--19\,533, 2020.

\bibitem[Lippmann et~al.(2000)Lippmann, Haines, Fried, Korba, and
  Das]{lippmann2000analysis}
R.~Lippmann, J.~W. Haines, D.~J. Fried, J.~Korba, and K.~Das, ``Analysis and
  results of the 1999 darpa off-line intrusion detection evaluation,'' in
  \emph{International Workshop on Recent Advances in Intrusion
  Detection}.\hskip 1em plus 0.5em minus 0.4em\relax Springer, 2000, pp.
  162--182.

\bibitem[Yu et~al.(2016)Yu, Kolar, and Gupta]{yu2016statistical}
M.~Yu, M.~Kolar, and V.~Gupta, ``Statistical inference for pairwise graphical
  models using score matching,'' \emph{Advances in Neural Information
  Processing Systems (NeurIPS)}, vol.~29, 2016.

\end{thebibliography}

\vfill

\end{document}